%% file: ms.tex
\documentclass{article}

\usepackage{arxiv}

\usepackage[utf8]{inputenc} 
\usepackage[T1]{fontenc}    
\usepackage{hyperref}       
\usepackage{url}            
\usepackage{booktabs}       
\usepackage{amsfonts}       
\usepackage{nicefrac}       

\usepackage{graphicx}
\usepackage{cite}
\usepackage{amsmath}

\usepackage{multirow} 

\title{Naphthalene crystal shape prediction from molecular dynamics simulations}

\author{
  \textbf{Zoran Bjelobrk} \\
  Institute of Process Engineering, ETH Z\"{u}rich, CH-8092, Switzerland \\ \\
  \textbf{Pablo M. Piaggi} \\
  Theory and Simulation of Materials (THEOS), \'{E}cole Polytechnique \\ F\'{e}d\'{e}rale de Lausanne, CH-1015, Switzerland \\ 
  and Facolt\`{a} di Informatica, Istituto di Scienze Computationali, Universit\`{a} della \\ Svizzera italiana, via Giuseppe Buffi 13, CH-6900 Lugano, Switzerland \\ \\
  \textbf{Thilo Weber} \\
  Institute of Process Engineering, ETH Z\"{u}rich, CH-8092, Switzerland \\ \\
  \textbf{Tarak Karmakar} \\
  Department of Chemistry and Applied Biosciences, ETH Z\"{u}rich, \\ c/o USI Campus, via Giuseppe Buffi 13, CH-6900 Lugano, Switzerland \\
  and Facolt\`{a} di Informatica, Istituto di Scienze Computationali, Universit\`{a} della \\ Svizzera italiana, via Giuseppe Buffi 13, CH-6900 Lugano, Switzerland \\ \\
  \textbf{Marco Mazzotti} \\
  Institute of Process Engineering, ETH Z\"{u}rich, CH-8092, Switzerland \\ \\
  \textbf{Michele Parrinello} \\
  Department of Chemistry and Applied Biosciences, ETH Z\"{u}rich, \\ c/o USI Campus, via Giuseppe Buffi 13, CH-6900 Lugano, Switzerland \\
  and Facolt\`{a} di Informatica, Istituto di Scienze Computationali, Universit\`{a} della \\ Svizzera italiana, via Giuseppe Buffi 13, CH-6900 Lugano, Switzerland
}

\begin{document}
\maketitle

\begin{abstract}
We used molecular dynamics simulations to predict the steady state crystal shape of naphthalene grown from ethanol solution. 
The simulations were performed at constant supersaturation by utilizing a recently proposed algorithm [Perego \textit{et al.} J. Chem. Phys., 142, 2015, 144113].
To bring the crystal growth within the timescale of a molecular dynamics simulation we applied Well-Tempered Metadynamics with a spatially constrained collective variable, which focuses the sampling on the growing layer.
We estimated that the resulting steady state crystal shape corresponds to a rhombic prism, which is in line with experiments.
Further, we observed that at the investigated supersaturations, the $\{00\bar{1}\}$ face grows in a two step two dimensional nucleation mechanism while the considerably faster growing faces $\{1\bar{1}0\}$ and $\{20\bar{1}\}$ grow new layers with a one step two dimensional nucleation mechanism.
\end{abstract}

\section{Introduction}

Controlling the crystal shape of organic molecules is crucial for process optimization in a wide range of industries, in particular in the pharmaceutical industry. Active pharmaceutical ingredients (APIs) are commonly produced as crystal powders purified by crystallization from solution in stirred suspension crystallizers. The resulting crystal shape affects among other properties the filterability, tabletability, and bio-availability of the drug product. Hence, control of the crystal shape is crucial in the manufacturing of pharmaceutical compounds \cite{Winn2000, Li2016}. Experimental methods aimed at determining the crystal shape are often expensive and cumbersome.
For this reason optimizing the crystallization conditions to obtain the desired crystal shape can be a laborious trial-and-error endeavor. Obtaining insight into the growth process at the molecular level would be useful to make a rational choice for the crystallization conditions.

Molecular dynamics (MD) simulations have recently emerged as a useful tool to obtain atomistic insight into crystal growth \cite{Liu1995,Winn2002,Piana2005,Piana2006,Chen2010,Salvalaglio2012,Karmakar2018,Han2019}. In particular, the work of Salvalaglio \textit{et al.} \cite{Salvalaglio2013} showed the usefulness of MD simulations for the paradigmatic case of urea. 
In that work, the growth rate was calculated and the growth mechanisms, rough growth or two dimensional (2D) nucleation, were identified from simulations in which the crystal surface of interest was exposed to a supersaturated solution. Afterwards, the crystal shape was predicted based on the relative growth rates of the different crystal faces \cite{Chernov1962}.

The present work is dedicated to the study of naphthalene crystal growth in ethanol solution under different supersaturations\cite{Lovette2012,Grimbergen1998}. After urea, naphthalene is a step further towards the simulation of systems as complex as real life APIs.

The study of crystal growth of naphthalene using MD simulations presents at least two major challenges. First, the growth rates of the slowest growing crystal faces of naphthalene are so small, that growth of new layers is not observable within the timescale of a standard MD simulation ($\sim$ 1 $\mu$s). 
To overcome the simulation time limitation we employ Well-Tempered Metadynamics \cite{Barducci2008} (WTMetaD), which is a well established enhanced sampling method for the study of rare events.

Secondly, the strong finite size effects hamper direct comparison with experiments. For instance,  typical experiments are performed at approximately constant supersaturation while standard MD simulations are unable to maintain such conditions \cite{Salvalaglio2012, Salvalaglio2013}.
The finite size effects are addressed using the constant chemical potential molecular dynamics (C$\mu$MD) method developed by Perego \textit{et al.} \cite{Perego2015}, which maintains a constant supersaturation in the region adjacent to the crystal. With this simulation setup, relative growth rates of faces growing in the 2D nucleation regime can be obtained as a function of supersaturation.

\section{Methods}

\subsection{Chemical potential control}

\begin{figure*}
	\centering
	\includegraphics[width=\textwidth]{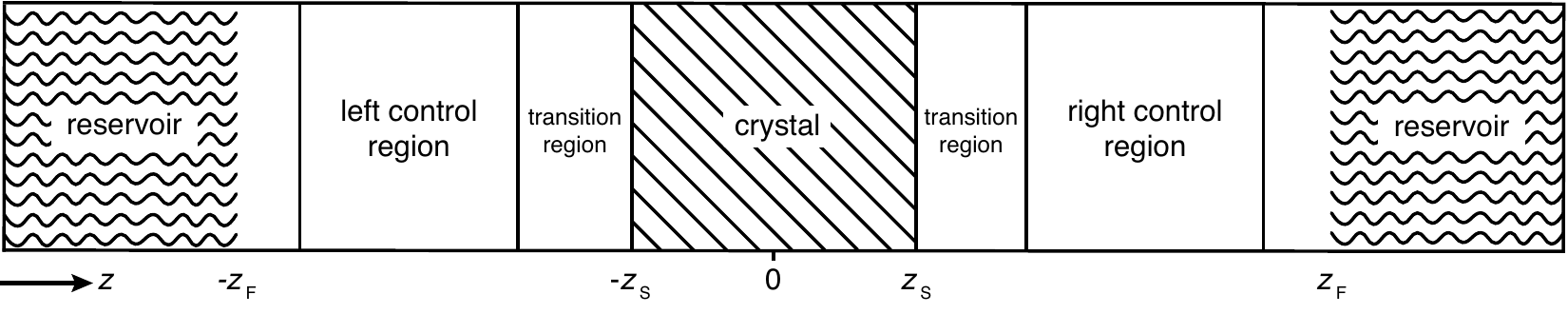}
	\caption{Schematic of the C$\mu$MD method illustrating the segmentation of the simulation box. The two crystal surfaces are prependicular to the $z$-axis and exposed to the solution at positions $-z_\text{S}$, and $z_\text{S}$ respectively. The external force positions are marked at $-z_\text{F}$ and $z_\text{F}$.} \label{fig:CmuMD_scheme}
\end{figure*}

In this section, we briefly summarize the C$\mu$MD method. Further details can be found in Reference \citenum{Perego2015}. The main features of this method are schematically depicted in Figure \ref{fig:CmuMD_scheme}: The system has periodic boundary conditions in all directions and is segmented along the $z$-axis into one reservoir, two control regions, as well as transition regions which surround the crystal slab. The slab is oriented such that the face of interest is exposed to the solution along the $z$-axis. 

The method works as follows: Two forces, $F_\text{L}^\mu$ and $F_\text{R}^\mu$, are introduced in order to control the flux of molecules between the reservoir and the left and right control regions, respectively. Each force acts along the $z$-axis at a distance $D_\text{F}$ from the crystal surface position $z_\text{S}$, namely at positions $-z_\text{F}$ and $z_\text{F}$, where $z_\text{F} = z_\text{S} + D_\text{F}$. The functional forms of the forces are
\begin{align}
F_\text{L}^\mu(t, z) &= k \left(C_{\text{CR},\text{L}}(t) - C_{0,\text{L}} \right) G_\omega(z, -z_\text{F})\\
F_\text{R}^\mu(t, z) &= k \left(C_{\text{CR},\text{R}}(t) - C_{0,\text{R}} \right) G_\omega(z, z_\text{F}),
\end{align}
where $t$ is the time, $C_{\text{CR},\text{L}}(t)$ and $C_{\text{CR},\text{R}}(t)$ denote the instantaneous solute concentrations, while $C_{0,\text{L}}$ and $C_{0,\text{R}}$ are the predefined set-values of the concentration in the left and right control regions respectively. $k$ is a force constant, and $G_\omega(z, z_\text{F})$ corresponds to the bell-shaped function defined as
\begin{equation}
G_\omega(z,z_\text{F}) = \frac{1}{4 \omega} \left[ 1 + \cosh \left( \frac{z - z_\text{F}}{\omega} \right) \right]^{-1},
\end{equation}
whose height and width are controlled by the parameter $\omega$.

The idea behind the method is that depending on the concentration in the control regions at time $t$, $F^\mu$	will accelerate the molecules towards or away from the control regions. In this way, the concentration in the control regions can be kept constant.

\subsection{Metadynamics}

The crystal growth rate depends on the growing crystallographic plane and the solution supersaturation. In the case of naphthalene, even at high supersaturations, the growth of certain faces takes place in timescales much longer than those that can be afforded using standard MD simulations. Here, we shall use WTMetaD, to bring the growth event within the timescale of the simulation. In this approach, a bias potential $V(\mathbf{s},t)$ that evolves in time is constructed as a function of a small number of collective variables (CVs) $\mathbf{s}$, which are functions of the atomic coordinates. The objective of $V(\mathbf{s},t)$ is to discourage frequently visited configurations and aid the system to surmount free energy barriers. $V(\mathbf{s},t)$ is defined as a sum of Gaussian functions of height $W$ and width $\sigma$.
$V(\mathbf{s},t)$ converges asymptotically to \cite{Valsson2016}
\begin{equation}
V(\textbf{s},t) = - \left( 1 - \frac{1}{\gamma} \right) F(\mathbf{s}) + c(t), \label{eq:V}
\end{equation}
where $F(\mathbf{s})$ is the free energy surface (FES); $\gamma$ is the so called bias factor, which takes a value greater than 1, and dictates how fast the height of the Gaussian functions decays with time. The time dependent energy offset $c(t)$ is independent on $\mathbf{s}$. 
The FES can be calculated as a function of a different set of CVs $\mathbf{s}'$ using the reweighting procedure decribed in Reference \citenum{Tiwary2015}. The interested reader is referred to References \citenum{Barducci2008} and \citenum{Valsson2016} for further details on WTMetaD.

\subsection{Surface layer crystallinity}

An important ingredient of WTMetaD is the choice of CVs. The CVs are typically non-linear functions of the atomic coordinates and describe the progress (state) of the process under study. In the case of crystallization, a reasonable choice is to use as CV the number of crystalline molecules in the growing layer. This CV would enhance both the growth and the dissolution of a single layer on the surface of the seed crystal. Here, we shall use a simple yet effective modification of the CV described in Reference \citenum{Salvalaglio2012} to localize the CV to the growing surface layer.

The CV should be able to account for the fact that molecules have a different environment in solution and in the crystal \cite{Santiso2011}. Molecules in solution are characterized by a small coordination number and almost random relative orientation of the neighbors. On the other hand, molecules in the crystal have higher coordination numbers and are aligned with respect to their neighbors. 
With this in mind, we define the surface layer crystallinity of solute molecule $i$ as
\begin{equation}
s_{\text{slc},i} = \phi_i \rho_i \Theta_i, \label{eq:s_c_i}
\end{equation}
where the function $\phi_i$ localizes the CV to the surface layer,
$\rho_i$ describes the local density, and $\Theta_i$ the local order. 

\begin{figure*}
	\centering
    \includegraphics[width=\textwidth]{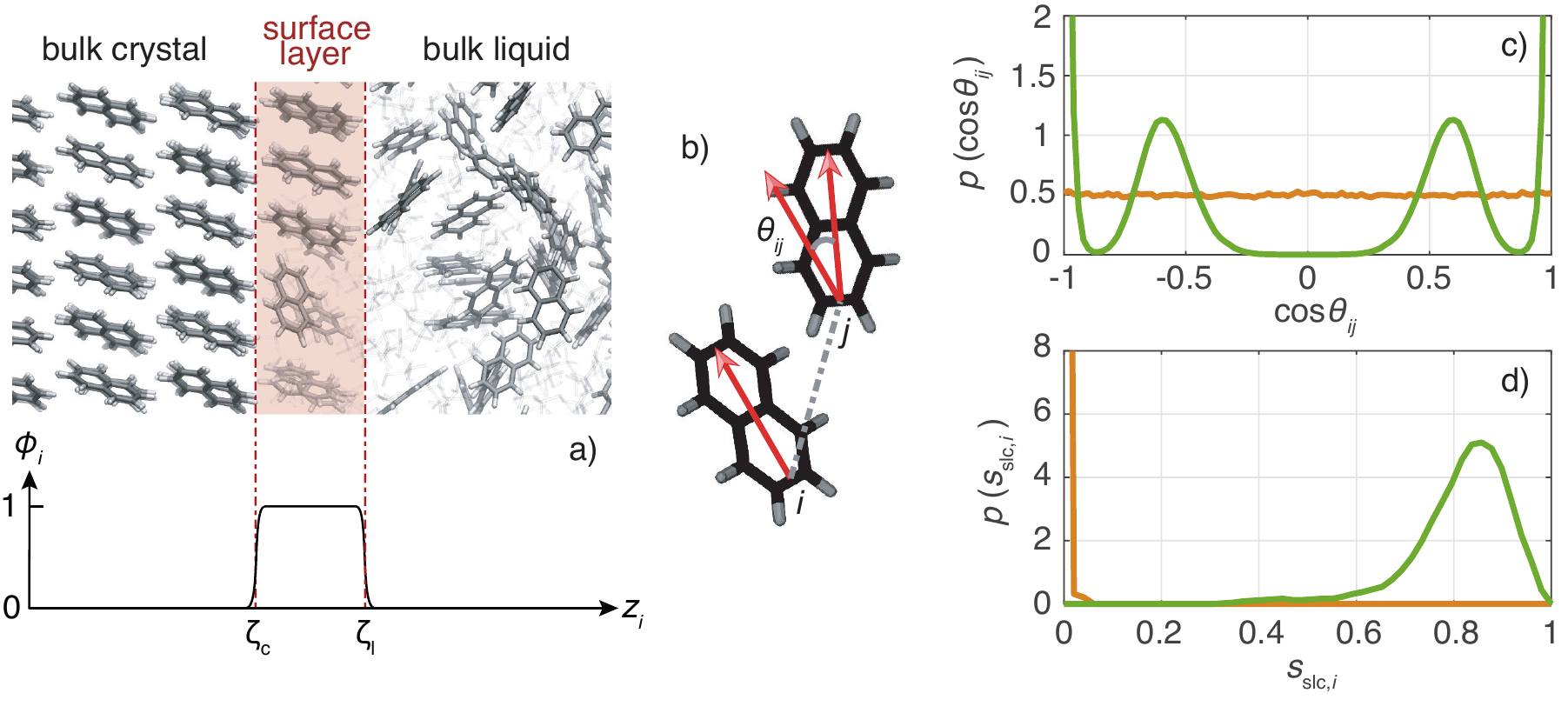}
\caption{a) Snapshot of a simulation box of the naphthalene $\{00\bar{1}\}$ face exposed to the solution (ethanol molecules are shown in faded colors). The red shaded region indicates the region of the growing surface layer. The surface layer crystallinity only takes molecules into account where function $\phi(z_i)$ is not zero (see graph below). b) Illustation of the vectors that characterize the orientation of naphthalene molecules $i$ and $j$ and the corresponding angle $\theta_{ij}$. c) Probability density function of $\cos \theta_{ij}$ for liquid (orange line) and crystalline (green line) naphthalene. d) Probability density functions of $s_{\text{slc},i}$ for each napthalene molecule $i$ in the case of a fully liquid surface layer (orange line) and fully crystalline surface layer (green line).}\label{fig:s_slc}
\end{figure*}

The CV $s_{\text{slc},i}$ needs to be continuous and differentiable. Function $\phi_i$ is defined for each solute molecule $i$ to have a non-zero value if solute molecule $i$ is located within the surface layer. Defining the $z$-axis as perpendicular to the crystal surface, $\phi_i$ takes the form
\begin{equation}
\phi_i = \left[ \frac{1}{1 + e^{-\sigma_c (z_i - \zeta_c)}} \right]
\left[ 1 - \frac{1}{1 + e^{-\sigma_l (z_i - \zeta_l)}} \right], \label{eq:phi}
\end{equation}
where $z_i$ is the position of molecule $i$ along the $z$-axis, $\zeta_c$ and $\zeta_l$ determine the boundaries of the surface layer, while $\sigma_c$ and $\sigma_l$ define the steepness of $\phi_i$ individually on the crystal and on the liquid side of the surface layer.
Figure \ref{fig:s_slc}a shows a snapshot of a simulation box of the $\{00\bar{1}\}$ face of naphthalene in which the surface layer is shaded in red. The $\phi_i$ as a function of $z_i$ for this configuration is shown below.

The local density $\rho_i$ is defined with the following rational switching function
\begin{equation}
\rho_i = \frac{ 1 - \left(\frac{ n_i }{ n_\text{cut} }\right)^{-a} }{ 1 - \left(\frac{ n_i }{ n_\text{cut} }\right)^{-2a}}, \label{eq:density}
\end{equation}
where $n_i$ is the coordination number of molecule $i$, $n_\text{cut}$ a predefined threshold value, and $a$ a parameter that determines the steepness of the switching function. If $n_i$ surpasses $n_\text{cut}$, $\rho_i$ tends to 1, whereas otherwise it tends to 0; $n_i$ is defined as
\begin{equation}
n_i = \sum_{\substack{j = 1 \\ j \neq i}}^N f_{ij},
\end{equation}
where $N$ is the number of solute molecules in the system and $f_{ij}$ is defined for solute molecule pairs $i$ and $j$ as
\begin{equation}
f_{ij} = \phi_j \frac{1 - \left(\frac{ r_{ij} }{ r_\text{cut} }\right)^{a} }{ 1 - \left(\frac{ r_{ij} }{ r_\text{cut} }\right)^{2a}}.
\end{equation}
Function $\phi_j$, which was introduced in Equation \eqref{eq:phi}, provides non-zero values only for neighboring molecules $j$ within the surface layer. The distance between the molecule pair is given by $r_{ij}$ and $r_\text{cut}$ is the cutoff for the calculation of the coordination number.

Now we turn to describe the part of $s_{\text{slc},i}$ that characterizes the local order. The orientation in space of each molecule is characterized by a vector $\textbf{\textit{l}}$. In the case of naphthalene, $\textbf{\textit{l}}$ is defined along the long axis of the molecule, see Figure \ref{fig:s_slc}b. The relative orientation between solute molecules $i$ and $j$ is characterized through an angle $\theta_{ij}$ defined as
$\cos \theta_{ij} = \textbf{\textit{l}}_i \textbf{\textit{l}}_j/(|\textbf{\textit{l}}_i| |\textbf{\textit{l}}_j|).$
In Figure \ref{fig:s_slc}c we show the probability density function of $\cos \theta_{ij}$ for neighbors $ij$ inside the cutoff $r_\text{cut}$ both for the liquid and the crystalline phase of naphthalene. The histogram shows four well defined peaks, each peak corresponding to one of the reference angles $\bar{\theta}_\kappa$ present in the crystal. We can now define the local order in Equation \eqref{eq:s_c_i} as
\begin{equation}
\Theta_i = \frac{1}{n_i} \sum_{\substack{j = 1 \\ j \neq i}}^N \left[ f_{ij} \sum_{\kappa = 1}^K  \exp \left\{- \frac{(\theta_{ij} - \bar{\theta}_\kappa)^2}{2 \sigma_\kappa^2} \right\} \right],\label{eq:Theta_i}
\end{equation}
where $\sigma_\kappa$ is the width of the angular distribution associated to $\bar{\theta}_\kappa$. $K$ denotes the total number of reference angles occurring in the crystal. The purpose of $f_{ij}$ in Equation \eqref{eq:Theta_i} is to restrict the summation on $j$ only to the neighbors within $r_\text{cut}$. Also, dividing by $n_i$ normalizes $\Theta_i$ in the interval $[0,1]$. Values of $\Theta_i$ close to 1 correspond to an environment where neighbors are aligned according to the reference angles $\bar{\theta}_\kappa$, while values of $\Theta_i$ close to 0 characterize environments with random orientations relative to their neighbors.

The combined effect of $\phi_i$, $\rho_i$, and $\Theta_i$ leads to a crystallinity measure $s_{\text{slc},i}$ that takes into account both density and order around the molecule $i$ within the surface layer. Taking the sum over all $N$ solute molecules of the system leads to the surface layer crystallinity
\begin{equation}
s_\text{slc} = \sum_{i = 1}^N s_{\text{slc},i}, \label{eq:s_c}
\end{equation}
which quantifies the crystallinity of the growing layer and avoids distortions of the bulk liquid as well as the bulk crystal in biased simulations by the use of the function $\phi_i$. In Figure \ref{fig:s_slc}d we show probability density functions (PDFs) of $s_{\text{slc},i}$ for a fully liquefied layer and for a fully grown crystal layer. It is clear from the figure that $s_{\text{slc},i}$ distinguishes well these two states.

\subsection{Biased collective variable}

\begin{figure}
	\centering
	\includegraphics[width=\textwidth]{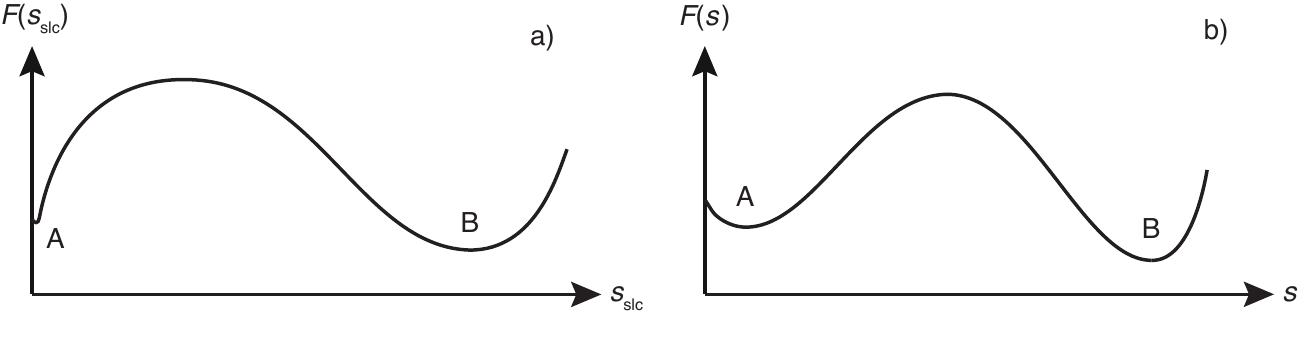}
	\caption{a) Scheme of a FES, $F$, in dependence of the surface layer crystallinity $s_\text{slc}$. The local minimum A is the fully dissolved and local minimum B the fully grown surface layer. b) Corresponding scheme of the FES in dependence of the biased CV, $s$, in which states A and B feature similar widths. \label{fig:FES_scheme}}
\end{figure}

Although $s_\text{slc}$ could have been used to construct the WTMetaD bias potential, we found it beneficial to use a slightly different CV. This is due to the fact that the two minima in the FES  corresponding to the fully dissolved layer (A) and fully grown layer (B) have rather different widths (see the illustrative scheme in Figure \ref{fig:FES_scheme}a). In particular basin A is rather narrow and close to the boundary of the CV. The first feature requires the use of very narrow Gaussians. These narrow Gaussians are totally unnecessary in the filling of basin B and thus convergence is slowed down. On the other hand, the closeness of basin A to the boundary creates artificial effects \cite{Micheletti2004,Crespo2010}.

In order to solve these problems, we could have used the variable Gaussian width approach of Branduardi \textit{et al.} \cite{Branduardi2012}. Here instead we prefer to use a much simpler fixup. We use as CV not directly $s_\text{slc}$ but the power law modification
\begin{equation}
s = s_\text{slc}^{\chi},
\end{equation}
where the exponent $\chi$ $<$ 1 is chosen to make the two minima approximately equally broad. A qualitative example of the effect of this change of CV is shown in Figure \ref{fig:FES_scheme}b. Not only are the two minima similarly broad but the minimum at A has been pushed fundamentally from the $s$ = 0 border. Adapting the CV in this way leads to an improved sampling of the low crystallinity states, while not affecting the sampling of high crystallinity states in an undesired manner.

The parameters used for $s_\text{slc}$, $s$, and WTMetaD are tabulated in the supporting information (SI).

\subsection{Force fields}

The general AMBER force field (GAFF) \cite{Wang2004,Wang2006} was used for naphthalene and ethanol. Both types of molecules have a fully atomistic description. For naphthalene, the relaxed structure and the restrained electrostatic potential (RESP) charges \cite{Bayly1993} were calculated with Gaussian 09 \cite{Gaussian09} using the density functional theory structural optimization routine performed at the B3LYP/6-31G(d,p) level. The partial charges of the naphthalene atoms were fitted to the RESP charges. The lengths of the bonds involving hydrogen have been fixed at the equilibrium value. Details on the force field parameters of naphthalene and the force field validation can be found in the SI\cite{daSilva2012,Cruickshank1957, Andrews1926}. Force field parameters of ethanol were taken from van der Spoel \textit{et al.} \cite{Spoel2012}.


\subsection{Simulation runs}

All simulations were performed with Gromacs 2016.5 \cite{Berendsen1995,Lindahl2001,Spoel2005,Hess2008b,Abraham2015} patched with a private version of Plumed 2.5 \cite{Tribello2014}. The temperature was set to 280 K in line with the experiments reported in Reference \citenum{Lovette2012} and was kept constant with the velocity rescaling thermostat \cite{Bussi2009}. Periodic boundary conditions were used. The particle mesh Ewald approach \cite{Darden1993} was employed for the long-range electrostatic interactions, and the LINCS algorithm \cite{Hess2008a,Hess2008b} was used to constrain the bonds involving hydrogens. The non-bonded cutoff for the van der Waals and electrostatic interactions was set to 1 nm.

\begin{figure*}[!h]
        \centering
        \includegraphics[width=\textwidth]{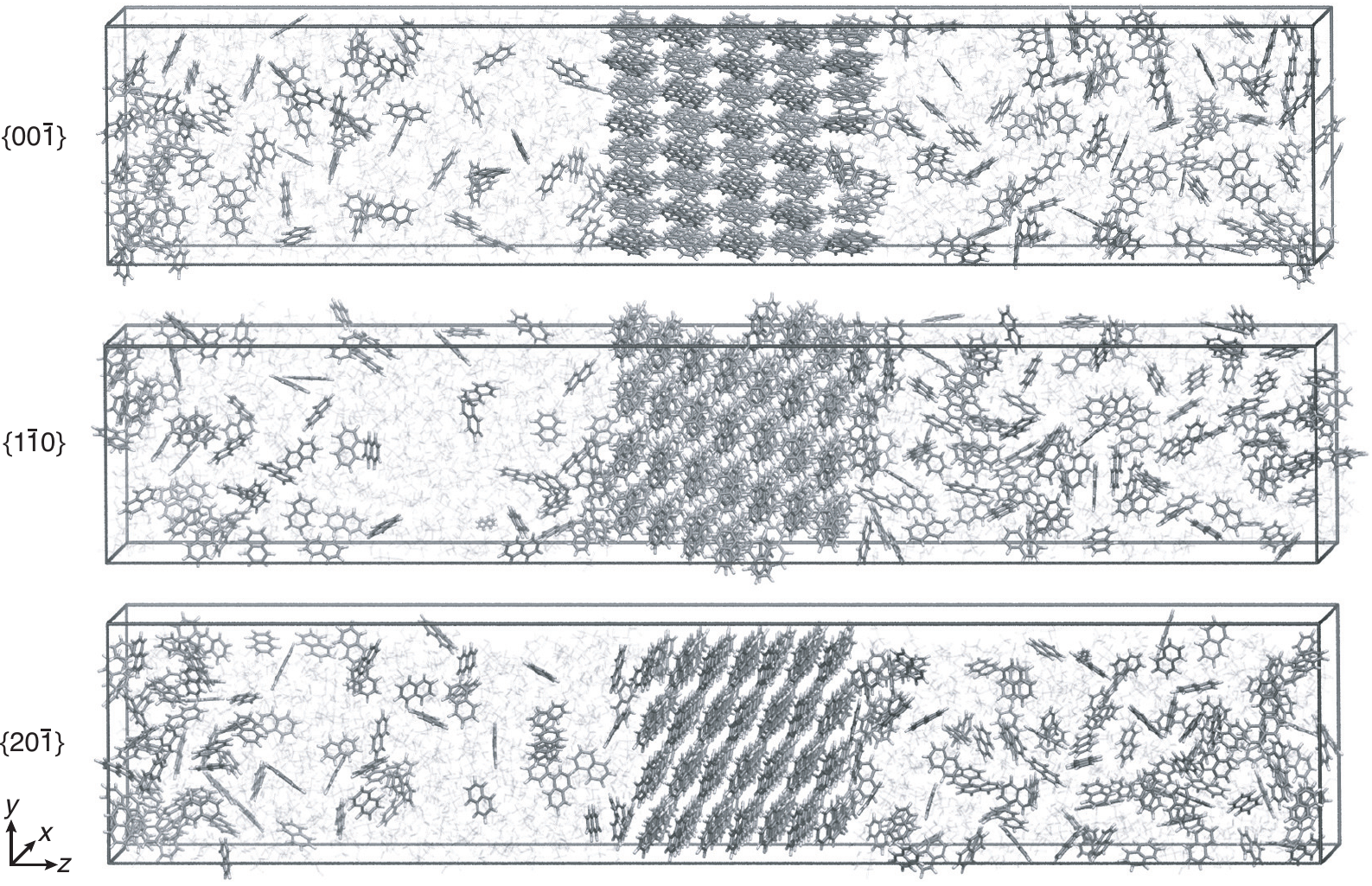}
        \caption{Simulation box visualization of the biased simulation setups for the three faces, $\{00\bar{1}\}$, $\{1\bar{1}0\}$, and $\{20\bar{1}\}$. Ethanol molecules are shown in faded colors.
        The visual molecular dynamics (VMD) software was used to visualize the trajectories and prepare the images\cite{Humphrey1996}.}\label{fig:Simulations}
\end{figure*}

For each of the three crystal faces, two concentrations of $C$ = 1.2 nm$^{-3}$ and $C$ = 1.5 nm$^{-3}$ were considered, which correspond to supersaturations of approximately $S$ $\approx$ 0.5 and $S$ $\approx$ 0.9 (see SI for details). The simulation boxes contained around 1300 molecules. 
The equilibration procedure\cite{Parrinello1981} to obtain the initial configuration is reported in the SI. To compute the free energy barrier, simulation runs of 1 $\mu$s each were performed using WTMetaD with $s$. Application of the C$\mu$MD algorithm ensured the constancy of the chosen concentration. Simulation box visualizations of the three faces, $\{00\bar{1}\}$, $\{1\bar{1}0\}$, and $\{20\bar{1}\}$, are shown in Figure \ref{fig:Simulations}.

During the course of the simulation, the crystal slab can experience a small random displacement. This can affect the determination of the position of the growing layer. To avoid this, the position of the center of mass of the two inner crystal layers was constrained using a harmonic potential. Further details can be found in the SI.

\section{Results}

\subsection{Steady state crystal shapes}

\begin{figure*}
	\centering
	\includegraphics[width=\textwidth]{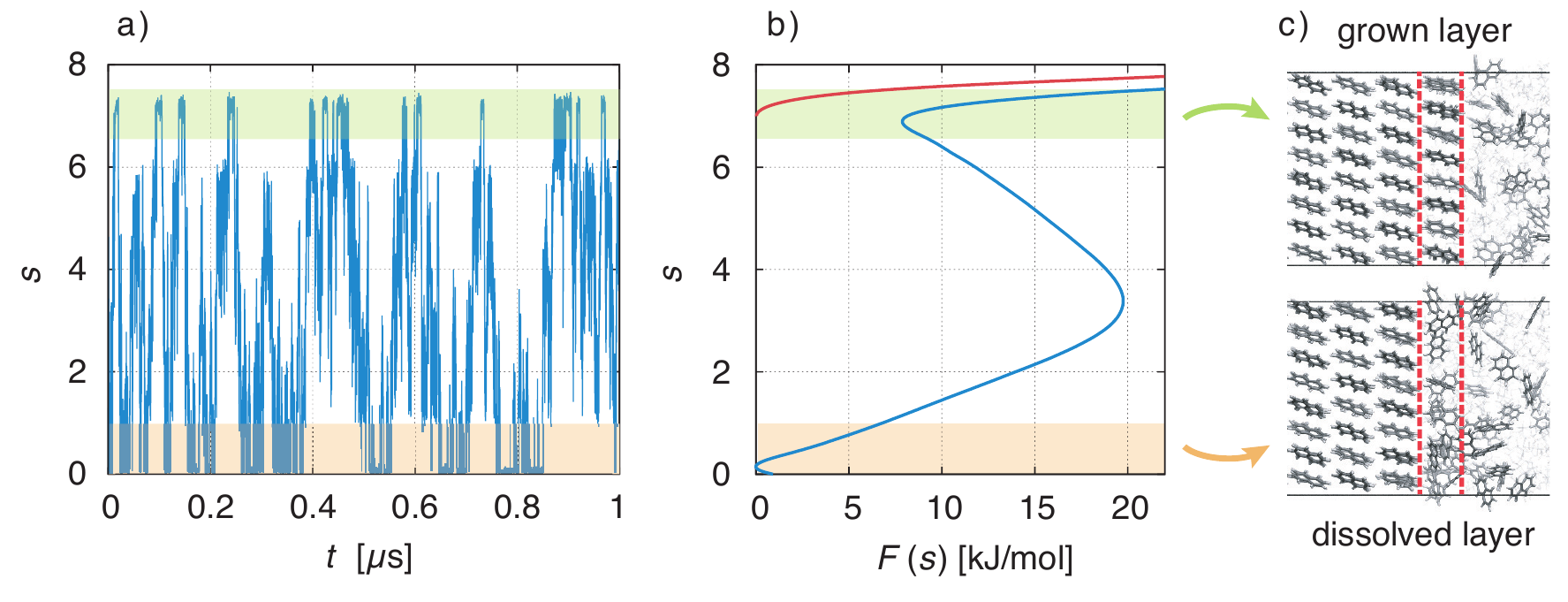}
	\caption{a) Time evolution of the biased CV, $s$. The orange shaded region relates to the dissolved crystal layer, while the green shaded region corresponds to the grown layer.
	b) Corresponding FES, $F(s)$ (blue curve), and wall potential (red curve).
	c) Representative simulation box snapshots of states at each minimum of the FES.  \label{fig:walker_sbias}}
\end{figure*}

The goal of this work is to predict the steady state crystal shape of naphthalene grown from an ethanol solution. The first step in this direction is to calculate the free energy barrier, $\Delta F_{\{hkl\}}^*$, for the growth of each face\cite{Salvalaglio2013}. We shall calculate $\Delta F_{\{hkl\}}^*$ from the WTMetaD simulations.

We describe a typical simulation taking as example the face $\{00\bar{1}\}$ at concentration $C$ = 1.5 nm$^{-3}$. In Figure \ref{fig:walker_sbias}a, we show the time evolution of the CV used in WTMetaD, namely $s$. During the simulation the system explores configurations in which the crystal layer is grown and dissolved. This should be compared to a standard MD simulation in which the system would remain for a much longer time in the dissolved state. From the simulation, the FES can be calculated using Equation \eqref{eq:V} and is shown in Figure \ref{fig:walker_sbias}b (blue curve). As expected, $F(s)$ shows two minima that correspond to the grown and dissolved states.
To improve sampling a wall potential was imposed to prevent the growth of the full layer. The wall potential is shown as a red curve in Figure \ref{fig:walker_sbias}b (see SI for details).
Representative simulation box snapshots for the two minima are shown in Figure \ref{fig:walker_sbias}c.

\begin{figure*}
	\centering
    \includegraphics[width=\textwidth]{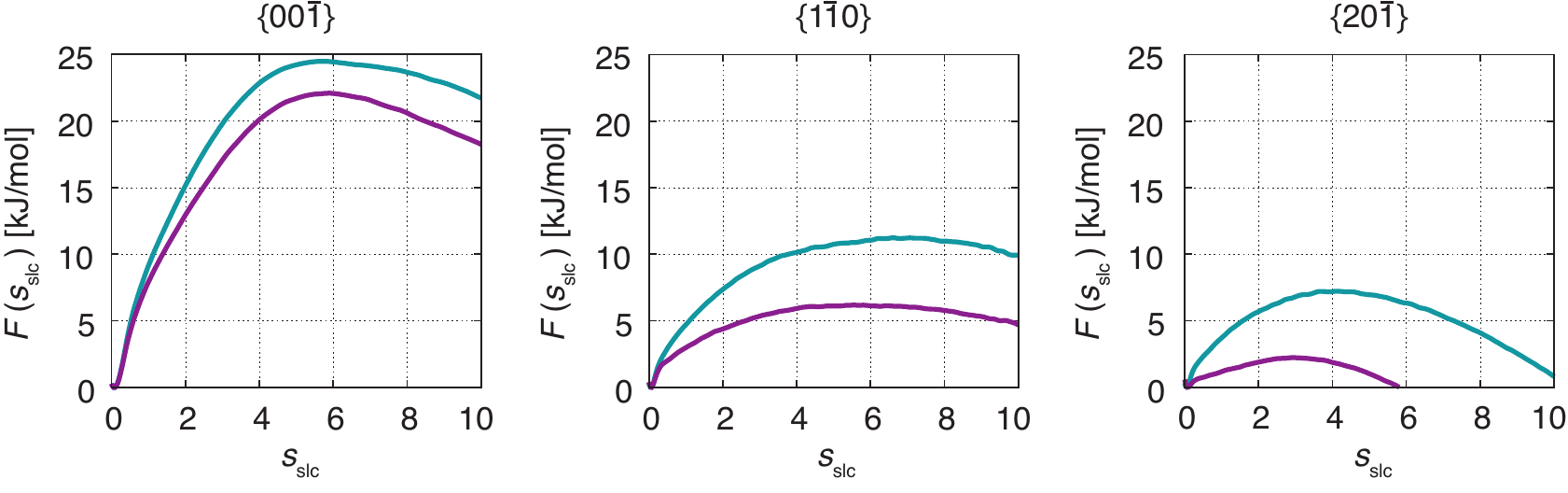}
	\caption{Free energy, $F(s_\text{slc})$, in dependence of crystallinity of the crystal surface, $s_\text{slc}$. Green denotes the concentration at $C$ = 1.2 nm$^{-3}$, and purple denotes the concentration at $C$ = 1.5 nm$^{-3}$.}\label{fig:FES}
\end{figure*}

Although $F(s)$ exhibits a free energy barrier, we decided to calculate the height of the barrier using a CV with a clear physical interpretation. For this reason, we reweighted\cite{Tiwary2015} our results to calculate the FES as a function of $s_\text{slc}$, which approximates the number of crystalline molecules in the growing layer. In Figure \ref{fig:FES}, we show $F(s_\text{slc})$ for all simulations, and the corresponding values of the $\Delta F_{\{hkl\}}^*$ are summarized in Table \ref{tab:F_star}. 
For both concentrations, $\Delta F_{\{00\bar{1}\}}^*$ is substantially larger than $\Delta F_{\{1\bar{1}0\}}^*$ and $\Delta F_{\{20\bar{1}\}}^*$. 
Furthermore, $\Delta F_{\{00\bar{1}\}}^*$ is less sensitive to changes in concentration than $\Delta F_{\{1\bar{1}0\}}^*$ and $\Delta F_{\{20\bar{1}\}}$. For face $\{00\bar{1}\}$, the free energy barriers differ between the two concentrations by $\sim$2.4 kJ/mol, while for faces $\{1\bar{1}0\}$ and $\{20\bar{1}\}$ the free energy barriers differ by $\sim$5.0 kJ/mol.

It is important to note, that the sampled $\Delta F_{\{hkl\}}^*$ are strongly affected by the finite size effects \cite{Salvalaglio2012}. In a small system, such as the one used in this work, the periodic boundary conditions of the simulation box induce an artificial stabilization of the emerging 2D nucleus on the surface of the crystal. This leads to an underestimation of the critical 2D nucleus size and as a consequence an underestimation of $\Delta F_{\{hkl\}}^*$. It is however possible to obtain trends which are consistent with experiments as shown in the following. Also one has to point out that here we are interested in the relative behaviors and we expect that finite size effects cancel out approximately.

\begin{center}
\begin{table}[h]
\small
  \caption{Free energy barriers (in [kJ/mol]).}
  \label{tab:F_star}
  \centering
  \begin{tabular*}{\textwidth}{@{\extracolsep{\fill}}llll}
    \hline
    & $\Delta F_{\{00\bar{1}\}}^*$ & $\Delta F_{\{1\bar{1}0\}}^*$ & $\Delta F_{\{20\bar{1}\}}^*$ \\
    \hline
	$C$ = 1.2 nm$^{-3}$		& 24.5		& 11.2		& 7.2	\\
	$C$ = 1.5 nm$^{-3}$		& 22.1		&	6.2		&	2.1	\\
    \hline
  \end{tabular*}
\end{table}
\end{center}

Now since we have the free energy barriers, we can compute the relative growth rates for each face and concentration. For the subsequent analysis we shall assume that all faces grow at the chosen supersaturations in the 2D nucleation regime. The perpendicular growth rate of a face $\{hkl\}$ growing by 2D nucleation can be written as \cite{Lovette2012,Liu1996}
\begin{equation}
v_{\{hkl\}} = \frac{d_{\{hkl\}}}{\tau_{\{hkl\}}},
\end{equation}
where $d_{\{hkl\}}$ is the interplanar spacing of the face and $\tau_{\{hkl\}}$ denotes the time to completely cover the crystal surface with a new layer of naphthalene molecules. Numerical values of $d_{\{hkl\}}$ are reported in the SI. $\tau_{\{hkl\}}$ depends on the nucleation rates and edge velocities of the emerging 2D nucleus. If we assume that nucleation is the rate limiting step, then we can approximate $\tau_{\{hkl\}}$ as the reciprocal of an Arrhenius type function \cite{Arrhenius1889,Liu1996,Salvalaglio2012}
\begin{equation}
\tau_{\{hkl\}}^{-1} = \xi \exp \left\lbrace \frac{-\Delta F^*_{ \{hkl\} } }{k_\text{B} T} \right\rbrace,\label{eq:Arrhenius}
\end{equation}
where $\xi$ is a pre-exponential coefficient which is assumed to be the same for all faces.

According to Chernov \cite{Chernov1962,Lovette2008}, the crystal morphology in a typical crystallization environment is determined by the crystal growth kinetics. The ratio of the perpendicular distances of faces $\{hkl\}$ and $\{h'k'l'\}$ from the center of mass of the crystal, $H_{\{hkl\}}$ and $H_{\{h'k'l'\}}$, is proportional to the ratio of the growth rates of the two faces, $v_{\{hkl\}}$ and $v_{\{h'k'l'\}}$
\begin{equation}
\frac{H_{\{hkl\}}}{H_{\{h^{\prime}k^{\prime}l^{\prime}\}}} 
\propto \frac{v_{\{hkl\}}}{v_{\{h^{\prime}k^{\prime}l^{\prime}\}}}.
\end{equation}
With the ratios $H_{\{hkl\}}/H_{\{h'k'l'\}}$, the steady state crystal shape can be estimated for each supersaturation by solving a system of plane equations in the Cartesian coordinates\cite{Zhang2006}.

\begin{figure}
	\centering
    \includegraphics[height=2.9cm]{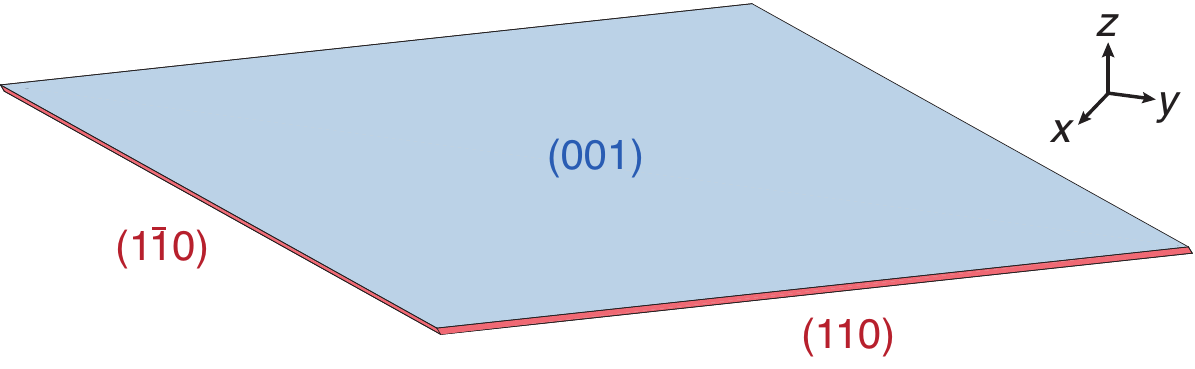}
	\caption{
	Steady state crystal shape obtained from simulation at $C$ = 1.2 nm$^{-3}$. The blue colored large surface corresponds to face $\{00\bar{1}\}$ and the red colored thin surfaces correspond to face $\{1\bar{1}0\}$. \label{fig:sscs}}
\end{figure}

For both concentrations, the predicted crystal shape is a wafer-thin rhombic prism, as shown in Figure \ref{fig:sscs} for $C$ = 1.2 nm$^{-3}$. The blue colored large surface corresponds to the slow growing $\{00\bar{1}\}$ face, while the red colored thin surfaces correspond to the fast growing $\{1\bar{1}0\}$ face. The very fast growing $\{20\bar{1}\}$ face is not present in the resulting steady state crystal shapes.

The activation energy barrier changes more rapidly with supersaturation for face $\{1\bar{1}0\}$ than for face $\{00\bar{1}\}$.
Consequently, the ratio $H_{\{1\bar{1}0\}}/H_{\{00\bar{1}\}}$ is about three times larger for $C$ = 1.5 nm$^{-3}$ than for $C$ = 1.2 nm$^{-3}$. Therefore the rhombic prism is about three times thinner for this higher concentration.

The simulation result at $C$ = 1.2 nm$^{-3}$ is comparable with the experiment at high supersaturation reported by Lovette \textit{et al.} in Reference \citenum{Lovette2012}.
Trends are captured well by the simulation, predicting that only faces $\{00\bar{1}\}$ and $\{1\bar{1}0\}$ are present and that the face $\{1\bar{1}0\}$ grows much faster in comparison to face $\{00\bar{1}\}$.

\subsection{Growth mechanisms}

\begin{figure}
    	\centering
       \includegraphics[width=14cm]{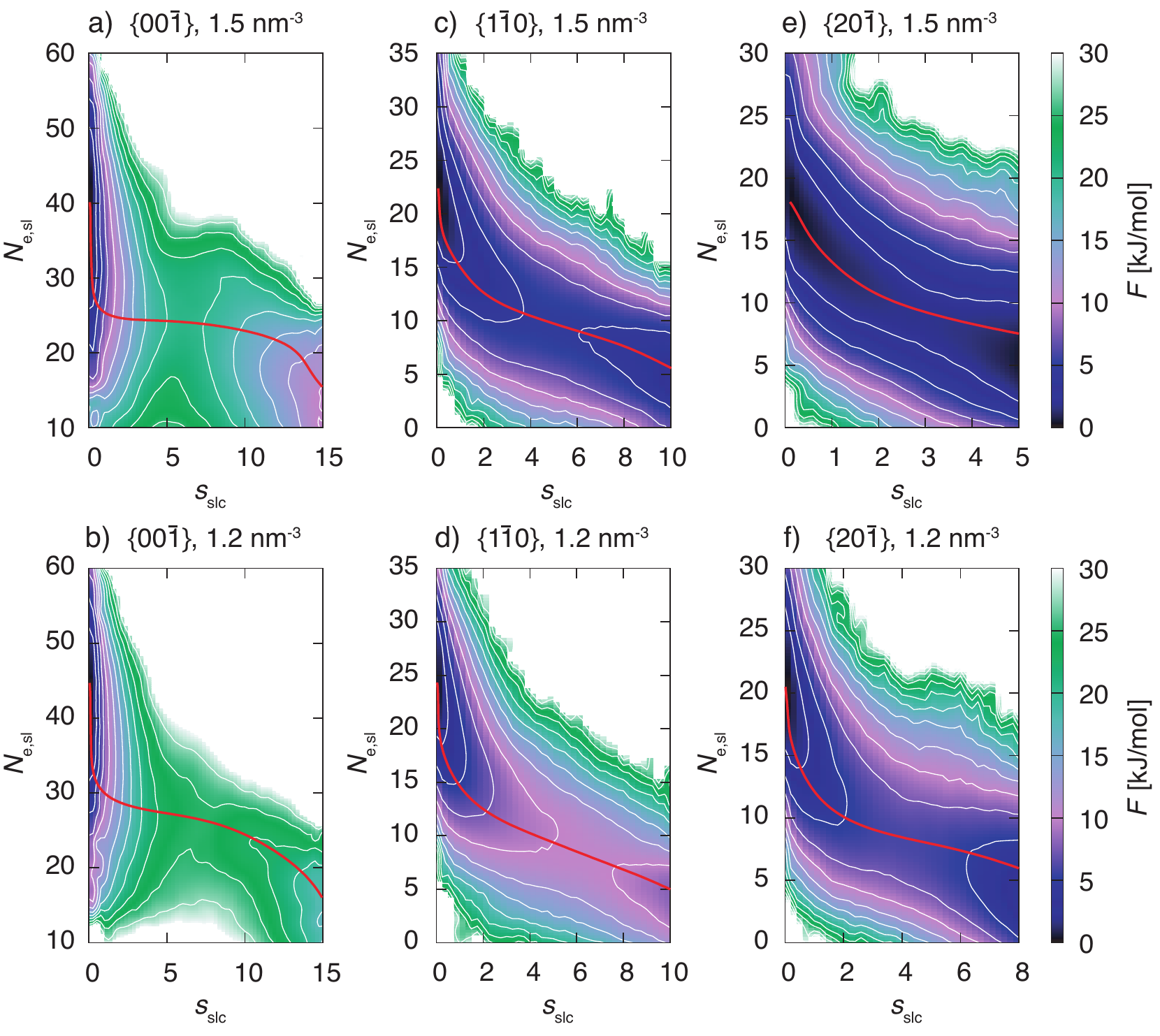}
	\caption{Free energy surfaces, $F$, in dependence of $s_\text{slc}$, the approximate number of crystalline napthalene molecules in the surface layer, and $N_\text{e,sl}$, the number of ethanol molecules in the surface layer. The white isocontours are drawn in 2.5 kJ/mol energy steps.
	The red lines correspond to the minimum free energy paths.
	}\label{fig:2D_e}
\end{figure}

The biased simulations also allow to gain insight into the details of the growth mechanism.
One should keep in mind that the simulations were performed with small surface areas exhibiting finite size effects. However, here again we are interested into trends which are expected to be the same for larger system sizes.
Figure \ref{fig:2D_e} shows the free energy surface, $F$, as a function of the CV $s_\text{slc}$, that approximates the number of crystalline naphthalene molecules in the surface layer, and CV $N_{\text{e},\text{sl}}$, i.e. the number of ethanol molecules residing in the surface layer.

We shall focus on the growth mechanism of face $\{00\bar{1}\}$ first (see for instance Figure \ref{fig:2D_e}a). The dissolved free energy minimum is strongly oriented along the $N_\text{e,sl}$ axis. Therefore the system will experience fluctuations in the number of ethanol molecules in the surface before starting to form a crystalline 2D nucleus. Of course, the number of naphthalene molecules in the surface is anticorrelated to the number of naphthalene molecules in the same region. Indeed, we observe in the simulation trajectories that an amorphous agglomerate of naphthalene molecules appears before the onset of crystallinity. Only at a subsequent stage the system surmounts the barrier in the $s_\text{slc}$ axis thus forming a crystalline 2D cluster that grows spontaneously.

From these observations we conclude that this face grows through a two step mechanism.
The first step is characterized by the formation of an amorphous agglomerate of naphthalene molecules while during the second step the molecules order into the crystalline lattice. This mechanism can be observed for both concentrations, see Figures \ref{fig:2D_e}a and \ref{fig:2D_e}b. The minimum free energy path for the transformation is presented in red in the FES and is compatible with the description given above.
The nudged elastic band method \cite{Henkelman2000} was used for the calculation of the minimum free energy paths.

Now we turn to describe faces $\{1\bar{1}0\}$ and $\{20\bar{1}\}$ that exhibit a growth mechanism different from the one of face $\{00\bar{1}\}$. We analyze in particular Figure \ref{fig:2D_e}c. In this case the decrease in the number of ethanol molecules in the surface occurs concomitantly with the increase in the number of crystalline naphthalene molecules. This is confirmed by the atomistic trajectories in which we also observe that as soon as naphthalene molecules reach the surface, they simultaneously tend to orient according to the crystal lattice. This description is compatible with a one step mechanism different from the scenario in face $\{00\bar{1}\}$. From the minimum free energy paths presented in red in Figure \ref{fig:2D_e} the two different growth mechanisms can be clearly distinguished.

\section{Conclusions}
In this work, we have used molecular dynamics to predict the steady state crystal shape of naphthalene grown from ethanol solution at two different supersaturations. We calculated the free energy barriers for the growth of faces $\{00\bar{1}\}$, $\{1\bar{1}0\}$, and $\{20\bar{1}\}$ using Well-Tempered Metadynamics. In order to mimic the experimental conditions the simulations were performed at constant supersaturation using the C$\mu$MD algorithm \cite{Perego2015}. From the free energy barriers we could estimate the relative growth rates and with them we predicted the steady state crystal shape. The resulting crystal shape is a wafer-thin rhombic prism, where only the $\{00\bar{1}\}$ and $\{1\bar{1}0\}$ faces are present and the former face grows considerably slower. These predictions are in line with experiments.

We identified two different growth mechanisms. For the $\{00\bar{1}\}$ face, a two step nucleation was observed in which an amorphous agglomerate of naphthalene molecules has to form before the molecules arrange into the crystalline lattice. For the $\{1\bar{1}0\}$ and $\{20\bar{1}\}$ faces the growth occurs with a concerted agglomeration and orientation of naphthalene molecules in the crystalline 2D nucleus.

The framework developed to enhance the growth of a single naphthalene crystal layer under constant supersaturation can be adapted to study the growth of more complicated organic molecules like APIs.

\section*{Conflicts of interest}
There are no conflicts to declare.

\section*{Acknowledgements}
Z.B. and M.M. are thankful for the financial support provided by Novartis Pharma AG, Lichtstrasse 35, 4056 Basel, Switzerland. 
P.M.P, T.K and M.P acknowledge support from the NCCR MARVEL funded by the Swiss National Science Foundation and from European Union Grant No. ERC-2014-AdG-670227/VARMET.
Z.B. would like to thank Matteo Salvalaglio, Dan Mendels, Claudio Perego, and Riccardo Capelli for valuable discussions. Z.B. thanks GiovanniMaria Piccini for supplying the code used to calculate the minimum free energy paths presented in Figure \ref{fig:2D_e}. The computational resources were provided by ETH Z\"urich and the Swiss Center for Scientific Computing at the Euler Cluster.





\bibliographystyle{unsrt}  
\bibliography{bibliography}

\newpage
\input{supplement.tex}

\end{document}

%% file: supplement.tex
\begin{center}
\vspace{5mm}
\LARGE{\textsc{Supplementary information}}
\vspace{2mm}
\end{center}

\section*{Parameters used for biased simulations}

Parameters of the surface layer crystallinity CV, $s_\text{slc}$, and biased CV, $s$, used in the simulations are summarized in Table \ref{tab:s_slc}. The Well-Tempered Metadynamics parameters are summarized in Table \ref{tab:MetaD_parameters}, where $\Delta s$ is the bin size of the grid on which the positions of the Gaussian potentials are discretized.

\begin{table}[!htb]
\small
\caption{Surface layer crystallinity CV parameters used for the biased simulations.}
\label{tab:s_slc}
\centering
\begin{tabular*}{\textwidth}{@{\extracolsep{\fill}}cccc}
\hline
										& $\{00\bar{1}\}$	& $\{1\bar{1}0\}$ & $\{20\bar{1}\}$	\\
\hline
$\bar{\theta}_1$ [rad]	& 0.0000					& 0.0000					& 0.0000					\\
$\bar{\theta}_2$ [rad]	& 0.4400					& 0.4400					& 0.4400					\\
$\bar{\theta}_3$ [rad]	& 2.7016					& 2.7016					& 2.7016					\\
$\bar{\theta}_4$ [rad]	& 3.1416					& 3.1416					& 3.1416					\\
$\sigma_1$ [rad]			& 0.15						& 0.15						& 0.15						\\
$\sigma_2$ [rad]			& 0.15						& 0.15						& 0.15						\\
$\sigma_3$ [rad]			& 0.15						& 0.15						& 0.15						\\
$\sigma_4$ [rad]			& 0.15						& 0.15						& 0.15						\\
$r_\text{cut}$ [nm]		& 0.65						& 0.90						& 0.90						\\
$n_\text{cut}$ [-]			& 4							& 4							& 4							\\
$a$ [-]								& 15							& 15							& 15							\\
$\zeta_c$ [-]					& 0.5884					& 0.5958					& 0.5958					\\
$\zeta_l$ [-]					& 0.6304					& 0.6229					& 0.6229					\\
$\sigma_c$ [-]				& 2000					& 2000					& 2000					\\
$\sigma_l$ [-]					& 2000					& 2000					& 2000					\\
$\chi$ [-]							& 0.7						& 0.5						& 0.5						\\
\hline
\end{tabular*}
\end{table}

\begin{table}[!h]
\small
\caption{Well-Tempered Metadynamics parameters used for the biased simulation.}
\label{tab:MetaD_parameters}
\centering
\begin{tabular*}{\textwidth}{@{\extracolsep{\fill}}cccc}
\hline
							& $\{00\bar{1}\}$	& $\{1\bar{1}0\}$ & $\{20\bar{1}\}$	\\
\hline
$W$ [kJ/mol]		& 0.8						& 0.5						&	0.5						\\
$\sigma_W$ [-]	& 0.7						& 0.4						&	0.4						\\
$\gamma$ [K]		& 12							& 10							&	8							\\
$\tau$ [ps]			& 1							& 1							&	1							\\
$\Delta s$ [-]		& 0.14						&	0.08						&	0.08						\\
\hline
\end{tabular*}
\end{table}

\section*{Naphthalene force field}\label{app:force_field}

\begin{table}[!h]
\caption{GAFF parameters for naphthalene.}
\label{tab:GAFF_parameters}
\centering
\begin{tabular*}{\textwidth}{@{\extracolsep{\fill}}ccccccc}
\hline
atom			& GAFF				& RESP				& mass			&	$x$		& $y$ 		& $z$		\\ 
name			& atom type		&  charge [e]	& [g/mol]		& [nm]		& [nm]		& [nm]		\\
\hline
C1				& ca					& -0.116476	& 12.010		& 0.000	& 0.244	& 	-0.071	\\
C2				& ca					& -0.116476	& 12.010		& 0.000	& 	0.244	& 0.071	\\
C3				& ca					& -0.264483	& 12.010		& 0.000	& 0.125	& 0.140 	\\
C4				& ca					& -0.264483	& 12.010		& 0.000	& 	-0.125	& 	0.140	\\
C5				& ca					& -0.116476	& 12.010		& 0.000	& -0.244	& 0.071	\\
C6				& ca					& -0.264483	& 12.010		& 0.000	& 0.125	& -0.140	\\
C7				& ca					& -0.116476	& 12.010		& 0.000	& -0.244	& -0.071	\\
C8				& ca					& 0.216029		& 12.010		& 0.000	& 0.000	& 0.072	\\
C9				& ca					& 0.216029		& 12.010		& 0.000	& 0.000	& -0.072	\\
C10				& ca					& -0.264483	& 12.010		& 0.000	& -0.125	& -0.140	\\
H1				& ha					& 	0.127069	& 1.008		& 0.000	& 0.338	& -0.125	\\
H2				& ha					& 	0.127069	& 1.008		& 0.000	& 0.338	& 0.125	\\
H3				& ha					& 0.145875		& 1.008		& 0.000	& 0.124	& -0.249	\\
H4				& ha					& 0.145875		& 1.008		& 0.000	& 0.124	& 0.249	\\
H5				& ha					& 0.145875		& 1.008		& 0.000	& -0.124	& 0.249	\\
H6				& ha					& 0.127069		& 1.008		& 0.000	& -0.338	& 0.125	\\
H7				& ha					& 0.127069		& 1.008		& 0.000	& -0.338	& -0.125	\\
H8				& ha					& 0.145875		& 1.008		& 0.000	& -0.124	& -0.249	\\
\hline
\end{tabular*}
\end{table}

\begin{figure}[!h]
        \centering
		\includegraphics[width=\textwidth]{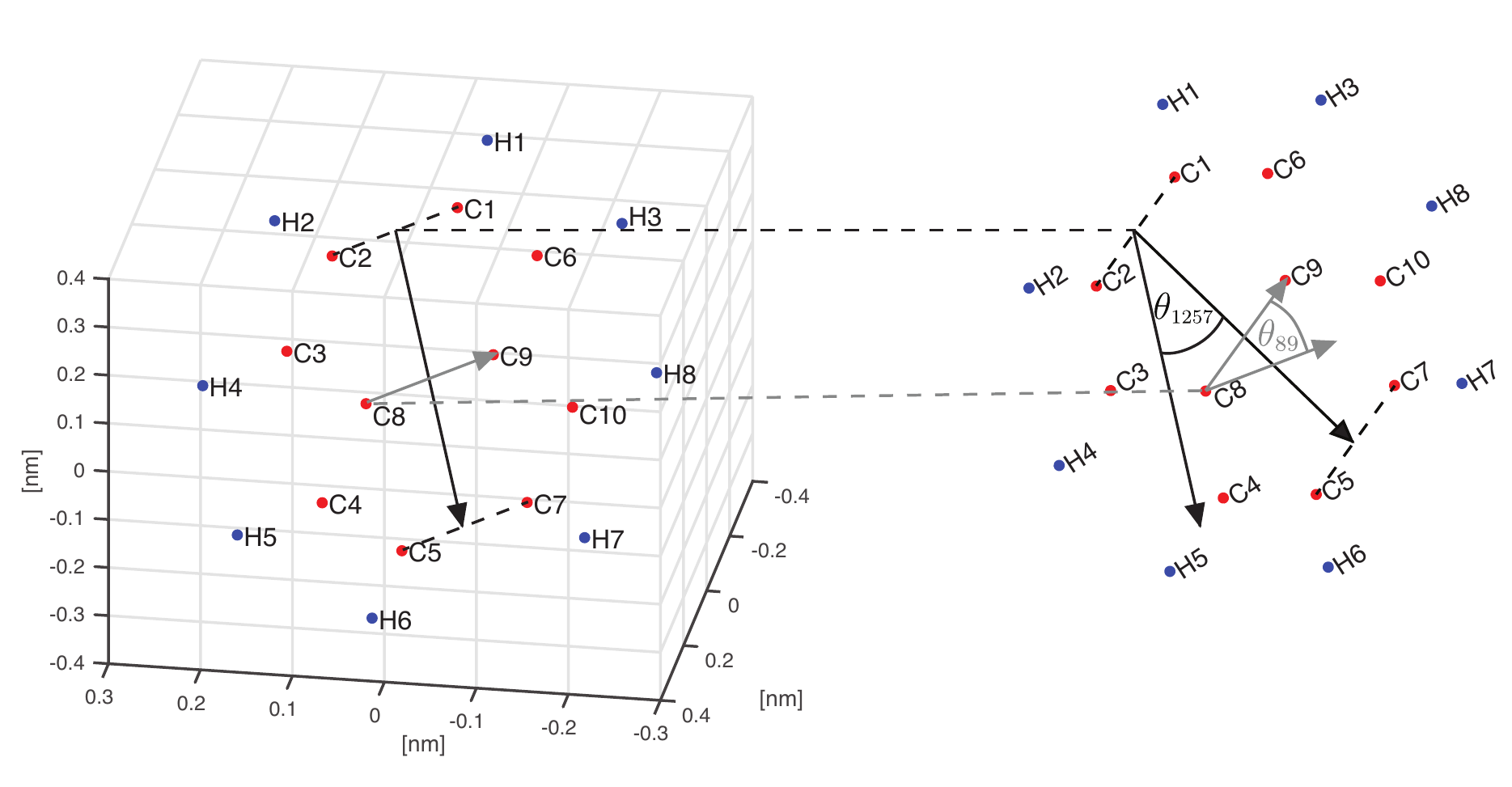}
        \caption{Naphthalene atom numbering and vector angle definitions of $\theta_{1257}$ and $\theta_{89}$.}\label{fig:naphthalene_def}
\end{figure}

The general AMBER force field (GAFF) \cite{Wang2004} parameters of naphthalene obtained by the procedure described in the main text are listed in Table \ref{tab:GAFF_parameters} (check Figure \ref{fig:naphthalene_def} for the numbering of the atoms).

\begin{figure}[!h]
	\centering
    \includegraphics[trim={0mm 0mm 0mm 0mm},clip,height=4.5cm]{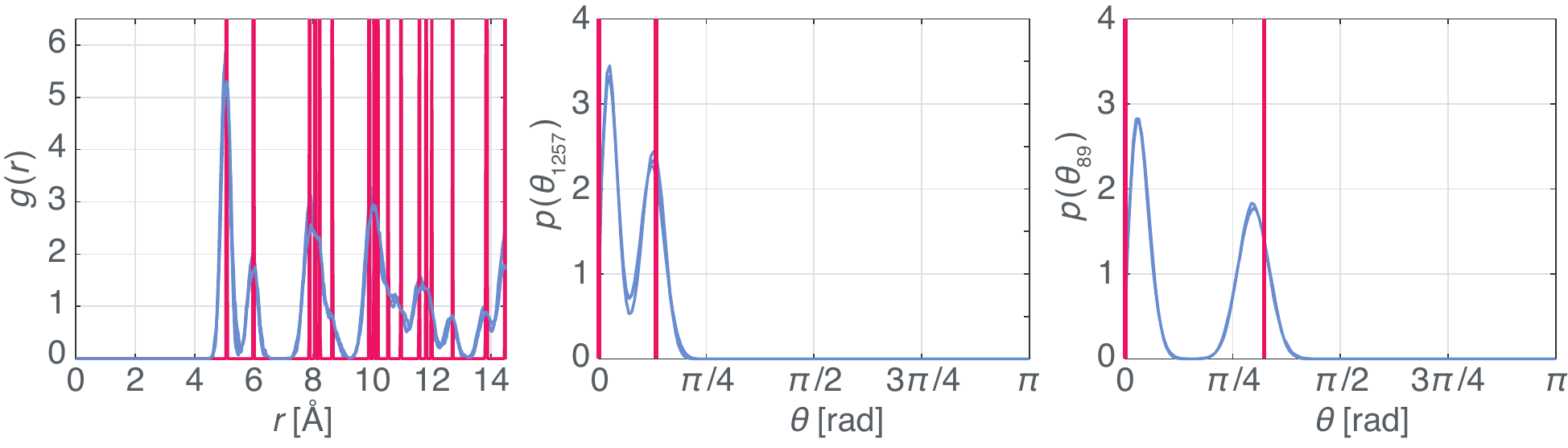}
    \caption{Histograms of the radial distribution function $g(r)$, left, and the vector angle probability density functions $p(\theta_{1257})$, center, and $p(\theta_{89})$, right. Histograms of the simulation (in blue) are shown at simulation times 3 ns, 6 ns, and 9 ns, and are in good agreement with the histograms obtained from the XRD measurements at 300 K (in red).}\label{fig:simulation_vs_XRD}
\end{figure}

To estimate how well the naphthalene GAFF reproduces the experimental crystal structure, simulations of a $\sim 3 \times 3 \times 3$ nm$^3$ crystal were performed under $NPT$ conditions with anisotropic pressure coupling \cite{Parrinello1981} at 1 bar and 300 K.
Figure \ref{fig:simulation_vs_XRD} shows the radial distribution function, $g(r)$, and the PDFs of the vector angles, $p(\theta_{1257})$ and $p(\theta_{89})$. The red lines were computed from XRD measurement data \cite{Cruickshank1957} and the three blue lines are simulation results at times 3 ns, 6 ns, and 9 ns. The center of mass of carbon atoms 8 and 9 were defined as molecule center and the vector angle definitions are shown in Figure \ref{fig:naphthalene_def}. Simulation results for $g(r)$, $p(\theta_{1257})$, and $p(\theta_{89})$ show all satisfactory agreement with the XRD data.

\begin{figure}[!h]
        \centering
		\includegraphics[trim={0mm 0mm 0mm 0mm},clip,height=8.5cm]{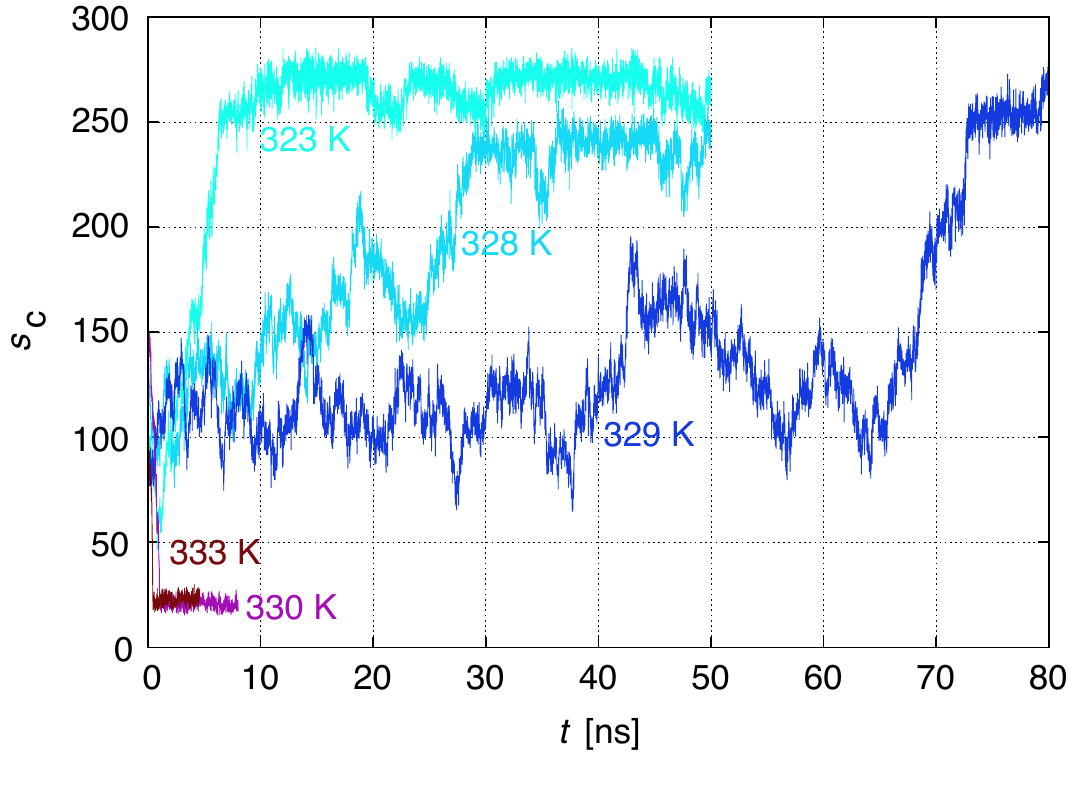}
        \caption{System crystallinity CV, $s_\text{c}$, in dependence of time for different temperatures. The seed crystal is associated with a $s_\text{c}$ value of $\sim 100$; For temperatures below 330 K, the crystalline phase grows until the liquid phase is completely converted into the crystalline state (at $s_\text{c}$ $\approx$ 270). At temperatures at and above 330 K, the crystal dissolves very quickly until only the liquid phase remains (at $s_\text{c}$ $\approx$ 20).}\label{fig:temperature}
\end{figure}

The melting temperature of the naphthalene GAFF was estimated by simulation runs of a roughly 3$\times$3$\times$3 nm$^3$ crystal slab exposing the $\{1\bar{1}0\}$ face to a liquid naphthalene phase around twice the size of the crystal.
Simulations were performed at different temperatures under $NPT$ conditions with the velocity rescale thermostat \cite{Bussi2009} and semi-isotropic Parrinello-Rahman barostat \cite{Parrinello1981} ($L_x$ and $L_y$ coupled, $L_z$ decoupled) at 1 bar. Growth of the crystal slab was observed at temperatures below 330 K, while at temperatures at and above 330 K the crystal slab was melting. The trajectories of the system crystallinity, $s_\text{c}$, for the different temperatures are presented in Figure \ref{fig:temperature}. The CV $s_\text{c}$ takes all solute molecules of the system into account and is reported in Reference \citenum{Salvalaglio2012}. The simulated melting temperature of $\sim 330$ K lies sufficiently close to the experimental one at 352 - 354 K \cite{Andrews1926}.

To obtain the initial configurations for both the crystallographic structure comparison and for the melting temperature, the equilibration procedure was used which is described in the following section. All MD simulation parameters which were utilized for the simulations above and are not explicitly mentioned, are described in the main text.

\section*{Simulation setup equilibration}

All equilibration steps were performed with Gromacs 2016.5 \cite{Abraham2015} using the velocity rescaling thermostat \cite{Bussi2009}, periodic boundary conditions, the particle mesh Ewald approach \cite{Darden1993} for the electrostatic interactions, and the LINCS algorithm \cite{Hess2008a,Hess2008b} to constrain the covalent bonds involving hydrogens. The non-bonded cutoff was set to 1 nm.

For each crystal face, the following procedure was performed to obtain initial configurations for the three slowest growing naphthalene crystal faces, $\{00\bar{1}\}$, $\{1\bar{1}0\}$, and $\{20\bar{1}\}$\cite{Grimbergen1998}. A seed crystal was constructed from the XRD data \cite{Cruickshank1957} with the face of interest perpendicular to the $z$-axis. The crystal system was first minimized with the conjugate gradient algorithm with a tolerance on the maximum force of 50 kJ mol$^{-1}$ nm$^{-1}$, and subsequently equilibrated under $NVT$ conditions for 1 ns, at 280 K and an integration time step of 0.5 fs. A temperature of 280 K was chosen to match the temperature of the experiments reported by Lovette \emph{et al.} \cite{Lovette2012}.

In a second step, the system was equilibrated for 10 ns under $NPT$ conditions with the anisotropic Parrinello-Rahman barostat \cite{Parrinello1981} at 1 bar, 280 K, and with the same integration time step of 0.5 fs. From the $NPT$ equilibration (while not considering the first 1 ns), the frame with box size values $L_x$, $L_y$, and the ratio $L_x/L_y$, closest to the average values, was considered. The averaged crystal configuration was then submerged in the solution by extending the box length $L_z$ to more than four times the crystal length in the $z$-direction and filling the box with naphthalene and ethanol molecules using the genbox utility of Gromacs \cite{Hess2008b}.

For the system containing the crystal as well as the solution, the same minimization and equilibration steps were performed. Although for the $NPT$ equilibration, the box expansion was only allowed in the $z$-direction, keeping $L_x$ and $L_y$ constant. The simulation frame with box length $L_z$ closest to the average one (again omitting the first 1 ns) was picked as initial configuration for the last equilibration step involving the C$\mu$MD algorithm \cite{Perego2015}.

To obtain the final initial configuration which holds the targeted concentrations in both control regions, the system had to undergo an equilibration procedure, in which the average concentration profile along the $z$-axis becomes static. A simulation of 25 ns ensured to reach an adequate convergence. The last frame of the concentration profile equilibration for each face and supersaturation was used for the simulations. The numerical parameter values of the C$\mu$MD algorithm are shown in Table \ref{tab:CmuMD_para}.

The number of molecules and simulation box lengths used for the biased simulations are presented in Table \ref{tab:box_properties}.

\begin{table}[!h]
\small
\caption{Parameters used for the C$\mu$MD algorithm. \label{tab:CmuMD_para}}
\centering
\begin{tabular*}{\textwidth}{@{\extracolsep{\fill}}cccccc}
\hline
$k$ [kJ/mol] & $\omega$ [-] & $z_\text{TR}$ [-] & $z_\text{CR}$ [-] & $z_\text{F}$ [-] & $\Delta z$ [-] \\
\hline
 1500 & 0.04 & 0.11 & 0.26 & 0.33 & 1/120 \\
\hline
\end{tabular*}
\end{table}

\begin{table}[!h]
\small
\label{tab:box_properties}
\caption{Number of molecules and simulation box lengths used for the biased simulations.}
\centering
\begin{tabular*}{\textwidth}{@{\extracolsep{\fill}}cccccc}
\hline
\multirow{2}{*}{face} & \multicolumn{2}{c}{number of molecules [-]} & \multicolumn{3}{c}{box lengths [nm]} \\
							& naphthalene	& ethanol		& $L_x$	& $L_y$	& $L_z$	\\
\hline
$\{20\bar{1}\}$	& 380				& 1053		& 2.975	& 3.577	& 16.560	\\
$\{1\bar{1}0\}$	& 360				& 951			& 3.035	& 3.172	& 16.936	\\
$\{00\bar{1}\}$ & 360				& 951			& 3.287	& 2.985	& 16.655	\\
\hline
\end{tabular*}
\end{table}

\section*{C$\boldsymbol{\mu}$MD concentration profiles}

\begin{figure}
	\centering
	\includegraphics[trim={0mm 0mm 0mm 0mm},clip,height=17cm]{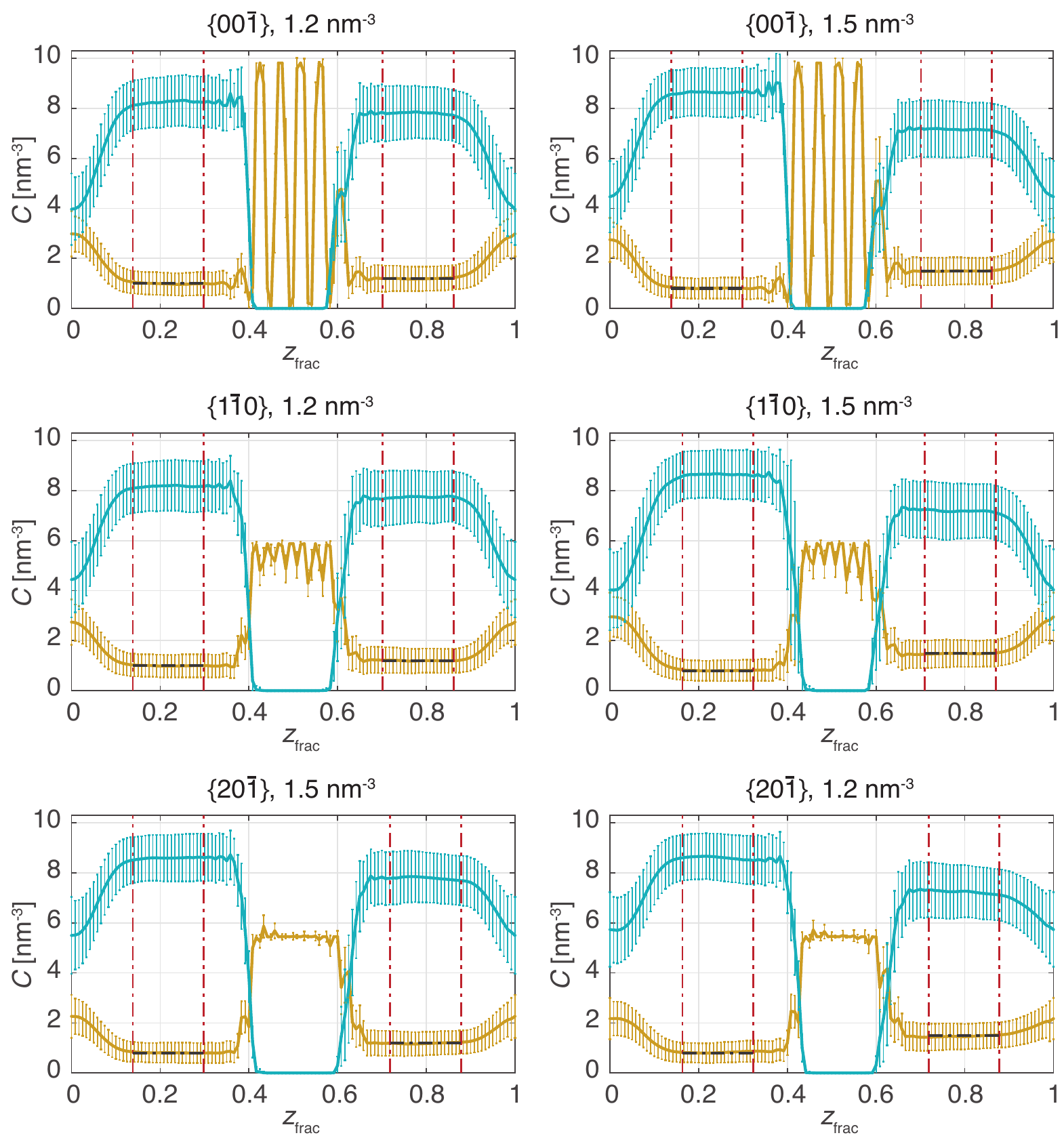}
	\caption{Concentration profiles along the $z$-axis (given as fractional coordinate) for the biased simulation averaged over 1 $\mu$s. The green lines correspond to the ethanol concentration and the khaki lines correspond to the naphthalene concentration. Error bars depict the standard deviation in the given segment. The red dashed lines indicate the boundaries of the control regions, in which the black dashed lines correspond to the target concentrations.}\label{fig:CmuMD}
\end{figure}

Parameters used in the C$\mu$MD algorithm are shown in Table \ref{tab:CmuMD_para}. And the corresponding concentration profiles for all six biased simulation runs (each performed for 1 $\mu$s) are shown in Figure \ref{fig:CmuMD}. The crystal surface was biased on the right side of the crystal shown in the middle of each graph, while the surface on the left was not biased at all. The target concentrations in both control regions were met with satisfactory accuracy.

\section*{Solubility estimation}

\begin{figure}
    \centering
	\includegraphics[trim={0mm 0mm 0mm 0mm},clip,height=6.2cm]{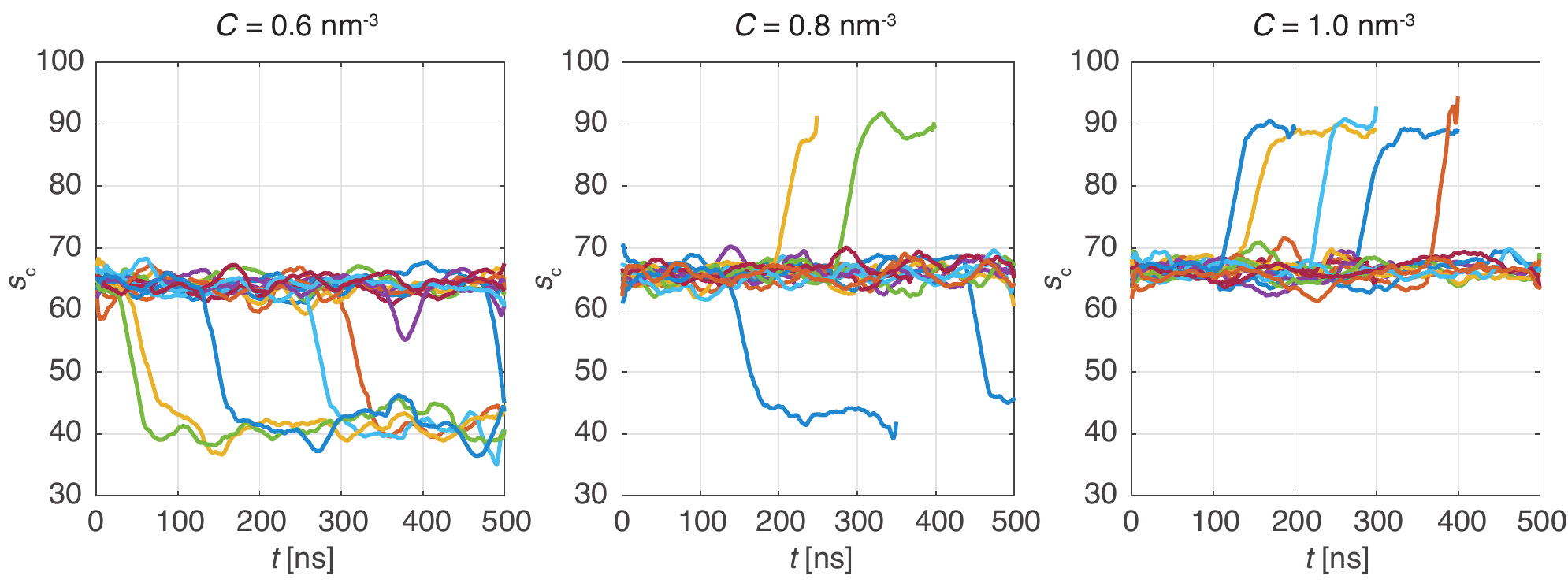}
	\caption{Unbiased C$\mu$MD simulation runs for three different naphthalene concentrations: system crystallinity, $s_\text{c}$, in dependence of time. The trajectories were smoothed out with a moving averge filter using a smoothing time of 3 ns to provide a clearer picuture of the growth and dissolution events. \label{fig:solubility}}
\end{figure}

To our knowledge, no simple algorithm exists for computing the exact solubility of naphthalene in ethanol at the given conditions. However it is possible to estimate the limits within the solubility is located by the use of the C$\mu$MD algorithm and long enough simulation runs. Figure \ref{fig:solubility} shows the system crystallinity CV, $s_\text{c}$, trajectories for different runs of the three simulation series, each under different concentrations, $C$ = 0.6 nm$^{-3}$, $C$ = 0.8 nm$^{-3}$, and $C$ = 1.0 nm$^{-3}$. For each concentration series the total time of all simulation runs sums up to 7.5 $\mu$s. In case of $C$ = 0.6 nm$^{-3}$, six dissolution events while no growth events were observed. For $C$ = 0.8 nm$^{-3}$ two dissolution and two growth events were observed. While for $C$ = 1.0 nm$^{-3}$, no dissolution, however five growth events were observed. The solubility is therefore estimated to reside around $C$ = 0.8 nm$^{-3}$.

\section*{Constraint of seed crystal movement}

For the biased simulations, the center of mass of the two most inner layers of each seed crystal was constrained in its movement along the $z$ axis with a harmonic potential. The force constant was set to 1500 kJ/mol for all three faces.

\section*{Use of wall potentials for biased simulations}

A lower and an upper wall potential were set to constrain the sampling to the relevant region:
\begin{equation}
V_\text{w} = \left\lbrace \begin{array}{cl}
w_\text{l} (s - s_\text{l})^2, &\text{if } s \leq s_\text{l}, \\
0, &\text{if } s_\text{l} < s \leq s_\text{u}, \\
w_\text{u} (s - s_\text{u})^2, & \text{else}.
\end{array} \right.
\end{equation}
The lower wall potential inhibits $s$ from getting stuck at the value of zero, causing the biased simulation to crash. While the upper wall plays the important role of not letting the system completing the biased layer, since it would increase the convergence time significantly. As already indicated by Piana \emph{et al.} \cite{Piana2005}, the removal of solvent molecules from the remaining surface gap of the almost fully grown layer corresponds to a slow process. The upper wall potential prevents the system from reaching the CV space where the surface gap would emerge.

\begin{table}[!h]
\caption{Wall potential parameters used for $s$.}
\label{tab:wall_s_bias}
\centering
\begin{tabular*}{\textwidth}{@{\extracolsep{\fill}}ccccccc}
\hline
			& \multicolumn{2}{c}{$\{00\bar{1}\}$} & \multicolumn{2}{c}{$\{1\bar{1}0\}$} & \multicolumn{2}{c}{$\{20\bar{1}\}$} \\
$C$ [nm$^{-3}$]			& 1.2	& 1.5	& 1.2	& 1.5	& 1.2	& 1.5	\\
\hline
$w_\text{l}$ [kJ/mol]	    & 75	&	75	& 50		& 50		& 50		& 50		\\
$s_\text{l}$ [-]				& 0.07	& 0.07	& 0.07	& 0.07	& 0.07	& 0.07	\\
$w_\text{u}$ [kJ/mol]	& 10		& 10		& 25		& 25		& 10		& 25		\\
$s_\text{u}$ [-]				& 7.0	& 7.0	& 3.5	& 3.5	& 3.5	& 3.0	\\
\hline
\end{tabular*}
\end{table}

\section*{Surface layer crystallinity trajectories}

\begin{figure}
        \centering
		\includegraphics[trim={0mm 0mm 0mm 0mm},clip,height=18cm]{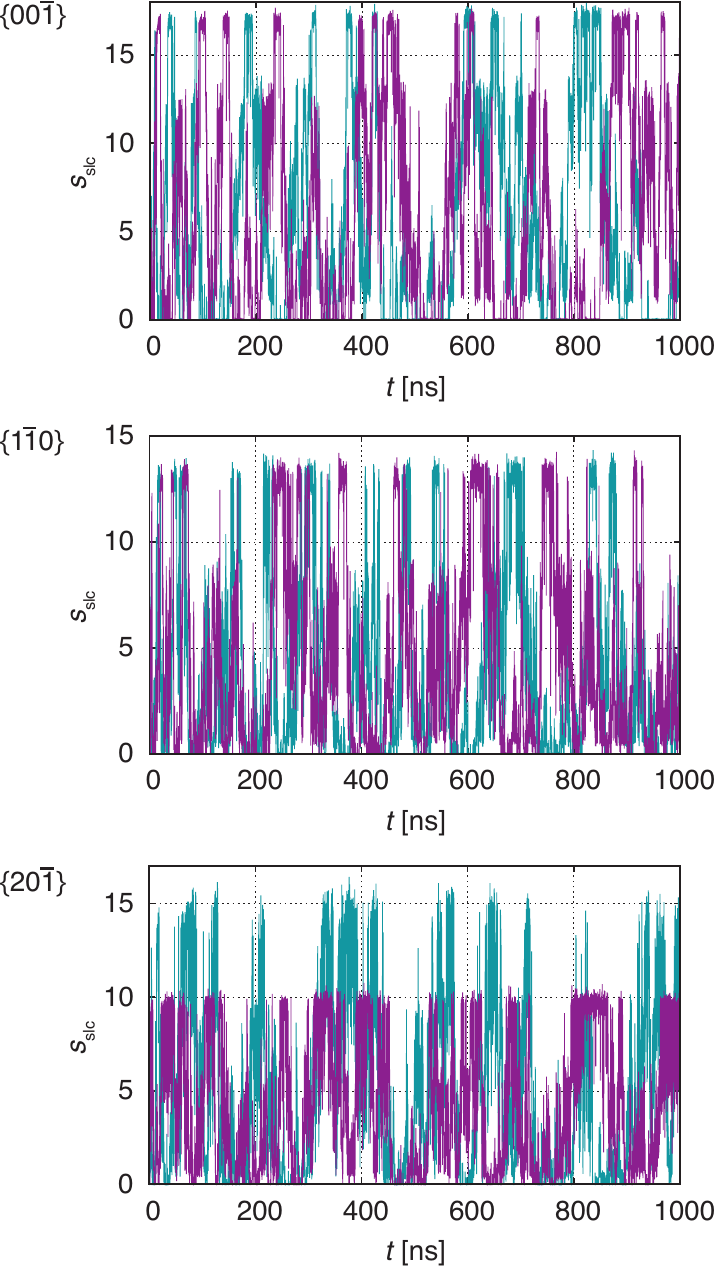}
        \caption{Trajectories of the surface layer crystallinity CV, $s_\text{slc}$, obtained from the biased simulations. Green denotes the concentration at $C$ = 1.2 nm$^{-3}$, and purple denotes the concentration at $C$ = 1.5 nm$^{-3}$.}\label{fig:walkers}
\end{figure}

The trajectories of the surface layer crystallinity CV, $s_\text{slc}$, of the biased simulations are shown in Figure \ref{fig:walkers}. The first 200 ns of $s_\text{slc}$ were not considered in the reweighting procedure \cite{Tiwary2015} for all reweighted plots reported in the main text.

\section*{Numerical values for the interplanar spacing}

Numerical values for the interplanar spacing, $d_{\{hkl\}}$, used for the calculation of the steady state crystal shape can be found in Table \ref{tab:d_and_S}. The values were obtained from $NPT$ simulations of a crystal of size $\sim 3 \times 3 \times 3$ nm$^3$ by using the anisotropic Parrinello-Rahman barostat \cite{Parrinello1981} and taking the average lengths of $d_{\{hkl\}}$. The $NPT$ simulations were performed for 10 ns each.

\begin{table}[!h]
\caption{Layer thickness, $d_{\{hkl\}}$, and simulation box surface area, $S_{\{hkl\}}$.}
\label{tab:d_and_S}
\centering
\begin{tabular*}{\textwidth}{@{\extracolsep{\fill}}cccc}
\hline
											& 	$\{00\bar{1}\}$ & 	$\{1\bar{1}0\}$	& $\{20\bar{1}\}$	\\
\hline
$d_{\{hkl\}}$ [nm]				& 	0.929					& 	0.451					& 	0.410	 				\\
\hline
\end{tabular*}
\end{table}